\documentclass[final,1p,times,number]{elsarticle}

\usepackage{amssymb,amsmath,amstext,lineno}                
\usepackage{graphicx}                                               
\usepackage{epstopdf}                                               
\usepackage{color}                                                     
\usepackage{bm}                                                        
\usepackage{appendix}                                              
\usepackage[utf8]{inputenc}
\usepackage{bbold}
\usepackage{bbm}
\usepackage[normalem]{ulem} 
\usepackage{latexsym}
\usepackage{xcolor}
\usepackage{braket}
\definecolor{lblue} {RGB}{51,71,158}
\usepackage[colorlinks=true,citecolor=blue,linkcolor=blue,urlcolor=lblue]{hyperref}

\modulolinenumbers[5]

\journal{Annals of Physics}

\begin{document}

\title{Nonergodic dynamics in disorder-free potentials}

\author{Ruixiao Yao}
\address{Department of Physics and State Key Laboratory of Low Dimensional
Quantum Physics, Tsinghua University, Beijing, 100084, China}

\author{Titas Chanda}
\address{Institute of Theoretical Physics, Jagiellonian University in Krak\'ow,  \L{}ojasiewicza 11, 30-348 Krak\'ow, Poland }

\author{Jakub Zakrzewski}
\ead{jakub.zakrzewski@uj.edu.pl}
\address{Institute of Theoretical Physics, Jagiellonian University in Krak\'ow,  \L{}ojasiewicza 11, 30-348 Krak\'ow, Poland }

\address{Mark Kac Complex
Systems Research Center, Jagiellonian University in Krakow, Krak\'ow,
Poland. }

\date{\today}

                              
\begin{abstract}
We review the dynamics of interacting particles in disorder-free potentials concentrating on a combination of a harmonic binding with a constant tilt. We show that a simple picture
of an effective local tilt describes a variety of cases. Our examples include spinless fermions (as modeled by Heisenberg spin chain in a magnetic field), spinful fermions
as well as bosons that enjoy a larger local on-site Hilbert space. We also discuss
the domain-wall dynamics that reveals nonergodic features even for a relatively weak tilt as suggested by Doggen et. al. [arXiv:2012.13722]. By adding a harmonic potential on top of the static field we confirm that the surprizing regular dynamics is not entirely due to  Hilbert space shuttering. It seems better explained by the inhibited transport within the domains of identically oriented spins. Once the spin-1/2 restrictions are
lifted as, e.g., for bosons, the dynamics involve stronger entanglement generation. Again for domain wall melting, the effect of the harmonic potential is shown to lead mainly to an effective local tilt.

\end{abstract}

\maketitle

\section{Introduction}
{The celebrated  contribution of Philip W. Anderson to the understanding of the relation between transport properties and disorder   \cite{Anderson58} is still providing the inspiration for current research. While Anderson model considered a single particle localization, the subject that occupied scientists for a number of years leading to the scaling theory of localization \cite{Abou73,Thouless74},  the effect of interactions has been frequently addressed too \cite{Anderson_MBL, nonperturbmbl}.  Only after almost 50 years, seminal }  works \cite{Gornyi05,Basko06} started a chain of many-body localization (MBL) studies resulting in hundreds of research papers partially summarized in recent reviews  \cite{Huse14, Nandkishore15, Alet18, Abanin19}.
In {short},  MBL is proposed as a robust counterexample to thermalization as localization is supposed to lead to the sustainable memory of the initial state in a closed system. Thus it is in a
violent contradiction with earlier advocated eigenstate thermalization hypothesis \cite{Deutsch91,Srednicki94}.  {MBL is characterized by the existence of a complete set of LIOMs (local integrals of motion) with exponentially vanishing local interactions, poissonian level statistics, eigenstates obeying the entanglement area law, a logarithmic growth of the entanglement entropy from initially weakly entangled or separable states and the already mentioned  memory effect of the initial states after quenches -- just to mention few main characteristics. Recently, however,}  \cite{Suntajs20e} put the very existence of MBL in the thermodynamic limit in question stimulating an intensive debate  \cite{Panda20,Sierant20b,Sierant20p,Suntajs20,Laflorencie20,Chandran21,Polkonikov20,Abanin21AP} based mainly on results coming from exact diagonalization studies. 
The second approach addresses time-dynamics with matrix product states techniques \cite{Znidaric08,Prosen09,Enss12,Bauer15,Hauschild16,Znidaric16,Enss17,Sierant17,Sierant18,Zakrzewski18,Doggen19,Doggen20,Chanda20t,Chanda20m}. The claims on the transition to MBL
proposed in these approaches typically mimics experimental approach \cite{Schreiber15,Luschen17} addressing the long time dependence of certain correlation functions. Still these claims are of limited value due to finite systems sizes and, importantly, rather short evolution times taken into consideration, times limited by the numerical methods based on matrix product states
related techniques such as time-evolving block decimation (TEBD) \cite{Vidal03,Vidal04,Znidaric08} or a more popular and recently developed time-dependent variational principle (TDVP) approach \cite{Schollwoeck11,Paeckel19}.

On the other hand, recent years brought several examples of strongly nonergodic behavior for specific initial states in disorder free models \cite{PRL_DFree} notably in situations where some conserved quantities exist. Subsequent studies linked such models with underlying gauge theories \cite{Smith17,Brenes18,Smith18,Chanda20,Magnifico20} and the corresponding conservation laws. On the other hand,  time dynamics of special initial states for Rydberg atom systems revealed  a nonergodic behavior  observed as pronounced long-time oscillations related to the so called scar states \cite{LukinRydScar, Turner18}. Nonergodic dynamics  is often being traced back to Hilbert-space fragmentation \cite{Khemani20, Sala20,Gromov20,Feldmeier20} resulting due to some almost conserved quantities.  Another possible source of disorder-free localization is the presence of the frustration in the system \cite{McClarty20}.

 The seemingly simple and prominent example of disorder-free localization appears in tilted lattices receiving the name of Stark many-body localization (SMBL) \cite{vanNieuwenburg19, Schulz19}. {It is an interaction extension of Wannier-Stark localization \cite{Emin87,Gluck02} occuring for noninteracting case. Here the localization comes from the fact that the kinetic energy (bounded by four times the hopping in tight-binding description) is comparably smaller than the potential difference between neighboring sites, resulting in a suppression of hopping and spreading  of particles. Clearly this picture should intuitively hold also for weak local interactions. SMBL was  
 first observed numerically \cite{vanNieuwenburg19, Schulz19} for one-dimensional (1D) spinless fermions case.   The experiments soon followed realizing tilted lattice models for spinful fermions in 1D \cite{Scherg20}, the two dimensional thermalization study \cite{Guardado20}, or in quantum simulator platforms  \cite{Guo20,Morong21}. This keen interest is due both to a rather straightforward implementation as compared to other
 interacting models without disorder mentioned above and to the desire to understand more deeply the origin of the nonergodic dynamics. Here some works stressed the importance of the global additional conservation laws (notably the dipole moment - see below), 
 that becomes valid, strictly speaking, in the large tilt limit only \cite{vanNieuwenburg19, Schulz19, Taylor20}. To avoid these effects a small harmonic tilt
 has been added \cite{vanNieuwenburg19,Taylor20} -- it has been also present in some experiments \cite{Scherg20}. A bit different understanding of the resulting nonergodicity has been advocated in 
 \cite{Guardado20}. Here the emergent dipole conservation is postulated when the potential energy associated with the tilt dominates the total energy. This interpretation has been further discussed in \cite{Khemani20}. The  failure of thermalization in such systems seems then to be direct consequence of energy conservation, thus one could retrieve the thermalization, which is slow but still present, by coupling to a bath \cite{Wu19} or in a higher dimensional system \cite{Guardado20}.} 

The aim of this work is to review and expand on a related but different situation, when both the tilt and an additional harmonic potential is present. Such an additional field may lead to phase separation in one-dimensional systems, where  localized and 
delocalized parts coexist in spatially separated regions as shown both for fermions and bosons    
 \cite{Chanda20c, Yao20b}. The phenomenon may be simply explained using the notion of a local tilt that enables to generalize these results 
 to other slowly varying-in-space potentials. The  global conservation laws for the ``charge'', ``dipole moment'' etc.  seem not very relevant for the dynamics on a realistic (i.e., experimentally feasible) time scales for typical initial states. We briefly discuss limitations of this approach seen, in particular, in the entanglement entropy dynamics.
 Recently the fate of specially prepared domain wall states has been studied \cite{Doggen20s} showing that their dynamics reveal strongly nonergodic behavior also for a relatively small tilt, in the regime, where level statistics \cite{Schulz19}  indicates ergodic behavior. Their results  
 were interpreted as a signature of ``shattered Hilbert space'', the term introduced in \cite{Khemani20}. We discuss physics of these states providing a discussion supplementing arguments provided in \cite{Guardado20,Khemani20}.
 
 Section 2 considers a Heisenberg spin chain in a combination of the constant tilt and the harmonic potential. Here ``local tilt'' is introduced
 and predictions obtained using this notion on the localization border are compared with spectral data \cite{vanNieuwenburg19,Schulz19}.
 The localization features for spinful fermions are discussed in Section~3, while bosons are briefly only considered in Section~4. The physics of domain walls is considered for spinless fermions and bosons in Section~5 and we conclude in Section~6.

\section{Heisenberg spin chain in a harmonic confinement}

Harmonic confinement is typical in cold atom studies although sometimes
 the effect  is  minimized \cite{Luschen17}. Quadratic dependence on the coordinate is quite common in quasi-1D situations realized in optical lattices
\cite{Fallani07,Zakrzewski09}, where a tight confinement in directions perpendicular to a chosen one is due to  strong laser beams with  Gaussian transverse profiles. Those profiles may be reasonably approximated as a harmonic trap along the considered direction \cite{Fallani07}. In recent works \cite{Chanda20c,Yao20b} we take a more brutal approach and assume a relatively
strong harmonic binding potential. We show that, surprizingly, such a harmonic binding may lead to a true separation of the space for one-dimensional interacting particles
with a central region being seemingly delocalized with the edges, on the other hand, being strongly localized.

Consider at first the spinless interacting fermions equivalent in 1D, via Jordan-Wigner transformation to interacting XXZ spin chain. Taking, as quite common in MBL studies,
the hopping and interaction strength constants to be equal we arrive at Hamiltonian:
\begin{equation}
 \label{eq: XXZ}
 H= J\sum_{l=-L/2}^{L/2-1} \vec{S}_l \ \cdot \vec{S}_{l+1} + \sum_{l=-L/2}^{L/2}  FlS^z_l
+ \frac{A}{2}\sum_{l=-L/2}^{L/2}l^2S^z_i,
\end{equation}
where $\vec{S}_l$ are 1/2-spins on different sites, $A$ is the curvature of the harmonic potential and $F$ is the magnitude of the possible additional tilt of the lattice. From now on we shall assume $J=1$ and take a unit lattice spacing. The model for $F=0$ and {very}  large $A$ obeys a quadrupole conservation law {(precisely in the $A\rightarrow\infty$ limit)},  while for $A=0$ and a very large $F$ the dipole conservation appears
\cite{Khemani20}. In these limits  the model belongs to {a class of} fracton systems where generically slow subdiffusive approach to thermalization is predicted \cite{Gromov20,Feldmeier20}. We shall restrict ourselves to moderate times of the order of thousands of hopping times and in advance we state that we did
not observe any traces of such slow processes in the model studied.

For a discussion of spectral properties of the system (e.g., the gap-ratio statistics) we refer the reader to \cite{Chanda20c}. Here, we visualize the time dynamics for moderate system size, $L=50$ easily accessible via time-dependent variational principle (TDVP) approach using matrix product states (MPS)  \cite{Haegeman11, Haegeman16, Paeckel19} . More specifically, we use a \textit{hybrid} variation of the TDVP scheme mentioned in \cite{Paeckel19, Goto19,Chanda20, Chanda20t}, where we first use two-site version of TDVP to dynamically grow the bond dimension up to a prescribed value, say $\chi_{max}$. After saturation occurs, we shift to the one-site version (avoiding  errors due to a truncation in singular values  in the two-site version \cite{Paeckel19, Goto19}). The final results are produced with $\chi_{max} = 512$. As the initial state here we take a N\'eel configuration with every second spin pointing up and every second down (along $z$-axis).

 \begin{figure}[h]
 \centering
 \includegraphics[width=0.9\linewidth]{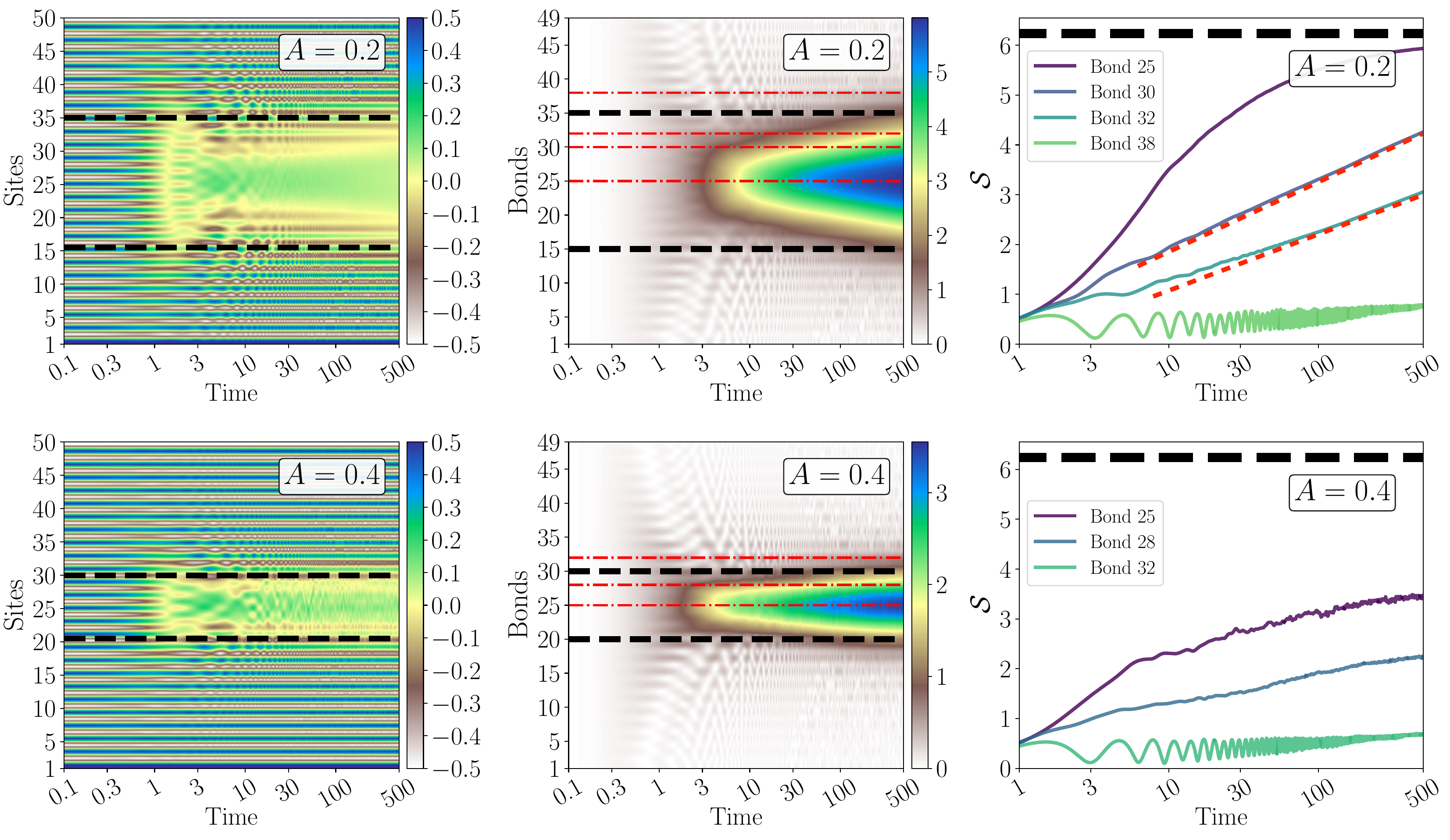}
  \caption{ {The Heisenberg chain in the harmonic trap.  Site-resolved spin dynamics, as measured by the local expectation values $\braket{S^z_l}$ for $L=50$ Heisenberg chain with no disorder -- left column for two values of the curvature $A$.  One clearly observes the coexistence of delocalized (in the center) and 
  localized regions (at the edges). The black dashed lines give the border of localization as given by Stark localization prediction with $F_c \approx 2$ {(for a discussion see text)}.
Middle column presents  the corresponding entanglement entropy dynamics across all bonds.
The entanglement grows rapidly in the central region remaining relatively low at the edges.
Right column: The entanglement entropy growth measured across selected bonds (marked by red dash-dot lines in the middle column). 
The black dashed lines show the maximum allowed value of $\ln 512$ of entanglement entropy by the MPS ansatz with $\chi_{max} = 512$.
    \label{dynheis} 
 }}
\end{figure}

Consider first the motion with no static field to observe the effects coming from a strong harmonic confinement. The time dynamics of the system is presented in Fig.~\ref{dynheis}
for two different values of harmonic potential curvature $A$. Observe that while at the center of the trap spin delocalize, the pattern of consecutive up and down spins is to a large extend preserved at both sides of the trap. This indicates that the system remembers the initial state in these regions  and thus is localized.

Bearing in mind the localization in the tilted lattice,  SMBL, the explanation of the observed phenomenon readily comes. For a site located at the distance $l_0$ from the center of the trap  the chemical potential is $\mu(l_0)=Al_0^2$ leading to the appearance of  the effective local static field 
$F = \frac{\partial}{\partial l_0} \left(\frac{A}{2} l_0^2\right) = l_0 A$. The localization length in the physical space for MBL and SMBL is quite small (otherwise the initial staggered pattern of spins would be washed out) so that a local tilt well describes a situation around $l_0$. 
When this local field exceeds the Stark localization border \cite{Schulz19, vanNieuwenburg19} -- the system behaves in this region as localized. The modulus of the local effective field increases towards the sides of the system so both the outer regions are localized  while the region close to the center remains extended. 
Note that one can always find localized regions for any finite values of $A$ for large enough systems under harmonic trapping potential - the situation not appearing for a pure Stark case.
The dashed lines in Fig.~\ref{dynheis} give the {apparent} Stark localization border $F \approx 2$.

 \begin{figure}[h]
 \centering
 \includegraphics[width=0.9\linewidth]{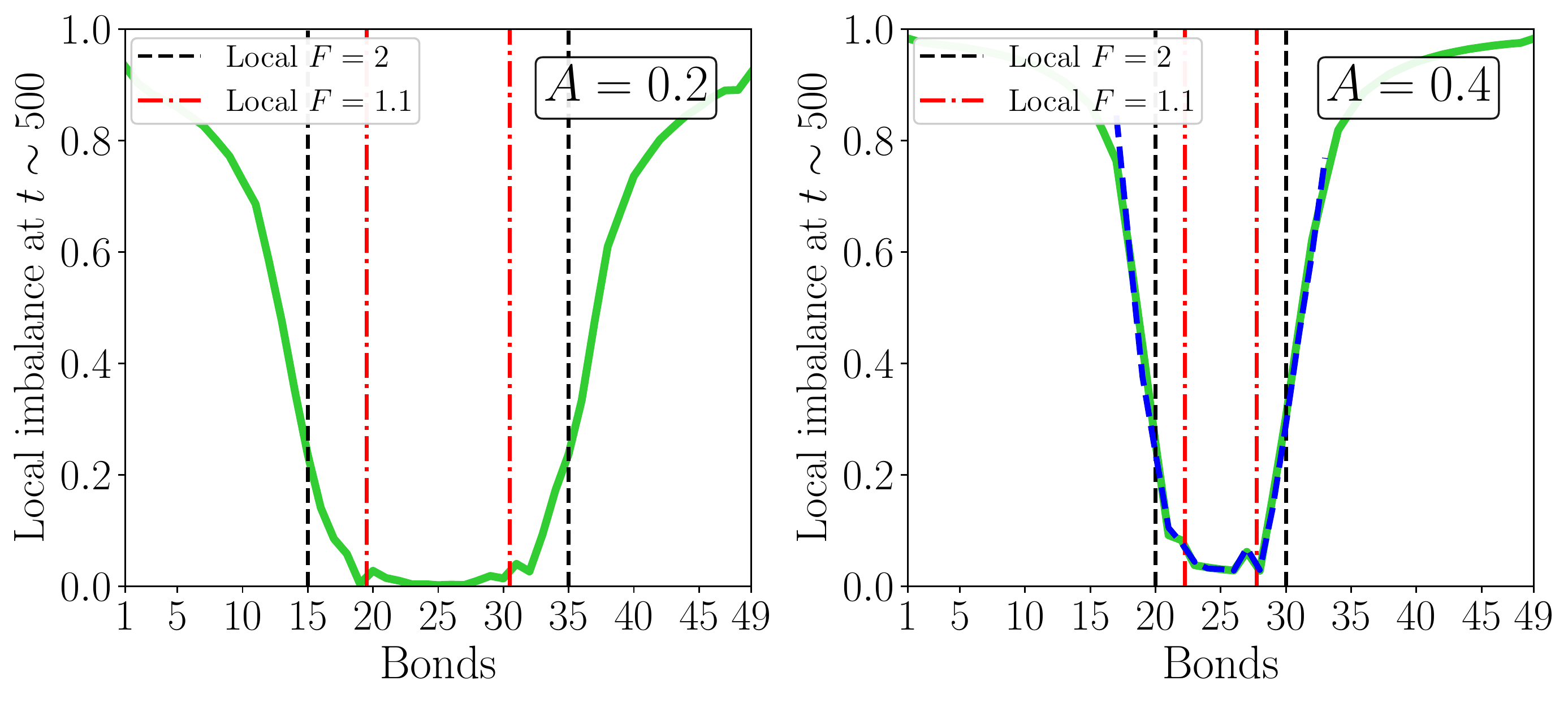}
  \caption{ {Green curves indicate long time local imbalance for  Heisenberg chain ($L=50$) dynamics in the harmonic trap for $A=0.2$ (left) and $A=0.4$ (right panel). 
Vertical black dashed lines correspond to a local field $F=2$ while red dash-dotted lines indicate position of local field $F=1.1$ corresponding to
the spectral critical value \cite{vanNieuwenburg19}. Clearly the spectral border matches points when imbalance becomes different from zero, while black lines indicating the visual border in the previous figure correspond to a significant local imbalance $I_l\approx 0.2$.
Blue dashed line give exact result for  a short $L=18$ chain indicating that finite size effects are very small in the case studied. 
    \label{dyn2} 
 }}
\end{figure}
{While this agreement was claimed in \cite{Chanda20c} one should observe the different  tilt  definition in  \cite{vanNieuwenburg19}!
In the units adopted here the critical value obtained from spectral statistics is actually $F_c^{corr}=1.1$. To understand the discrepancy
between the critical field value obtained from the spectral information and by us from time dynamics we analyze the final local imbalance
defined as $I(i)=2 |\langle S^z_i0)\rangle\langle S^z_i(t)\rangle + \langle S^z_{i+1}(0) \rangle\langle S^z_{i+1}(t)\rangle|$ as a function of bond position in the lattice. Since the data show residual oscillations we average this imbalance over time in $t\in[450,500]$ interval and plot it along the chain in Fig.~\ref{dyn2}. At the central delocalized regime the local imbalance vanishes then increases for a sufficiently large local tilt. Observe that the dashed black line
indicating $F=2$, the value taken as an indicative border in Fig.~\ref{dynheis} corresponds to a local imbalance value around 0.2. The
dash-dotted line indicate the position of the critical tilt value $F=1.1$ \cite{vanNieuwenburg19} which nicely coincides with positions where imbalance starts to be different from zero. Thus the local tilt description yields a good estimate of the transition point between localized and delocalized parts in the dynamics.}

{Note also that the blue dashed line in the right panel of Fig.~\ref{dyn2} which corresponds to a much smaller system size $L=18$ almost precisely matches the imbalance data for $L=50$. This seems to indicate that finite size effects  for disorder-free potentials are not significant. Since the finite region of transition between delocalized and localized regions seem to not depend on the system size, there is a ground to speculate that in our system this is not really a transition but a real crossover. This fact deserves a separate, more detailed study before drawing definite conclusions.}

At each bond linking sites $l$ and $l+1$ we may divide the system into two parts A and B. That allows us to define in a standard way the bond-dependent entanglement entropy $S_l= -\sum \lambda_i \ln \lambda_i$ where $\lambda_i$ are eigenvalues of the density matrix of one of the parts (A or B) obtained by tracing out the remaining part.  The bond dependent time dynamics of the entanglement entropy is presented in middle and right columns of Fig.~\ref{dynheis}. The rapid growth of the entropy at the center of the trap in delocalized regime allows us to expect that the results are not fully converged for $\chi_{max} = 512$. The growth of entropy slows down very fast when we move away from the center of the trap. 
In the outer regions the growth of entropy is very slow and even $\chi_{max} = 384$ (not shown) leads to fully converged results. Interestingly, the growth in the ``delocalized'' regime, close to the localization border seems logarithmic in time (compare  curves for bonds 30 and 32 for $A=0.2$). 
The possible explanation of this behavior comes from the fact that  entanglement entropies between neighboring sites cannot 
differ much as the local Hilbert space dimension is equal to 2. Thus logarithmic slow growth in time in localized region is ``exported'' to sites being the close neighborhood of the border between localized and delocalized sites.

One could wonder whether the observation for $L=50$ are sufficient to draw conclusions for larger system sizes, in the $L\rightarrow \infty$ limit. Observe that adding sites (on both sides symmetrically) to the system leads to an increase of regions with large value of the local electric field - the central region is not affected by this procedure. Thus contrary to the case of
MBL in random disorder where the thermodynamic limit is vividly discussed as mentioned in the introduction, in our case the large system size limit is almost trivial.

\begin{figure}[h]
\centering
\includegraphics[width=0.9\linewidth]{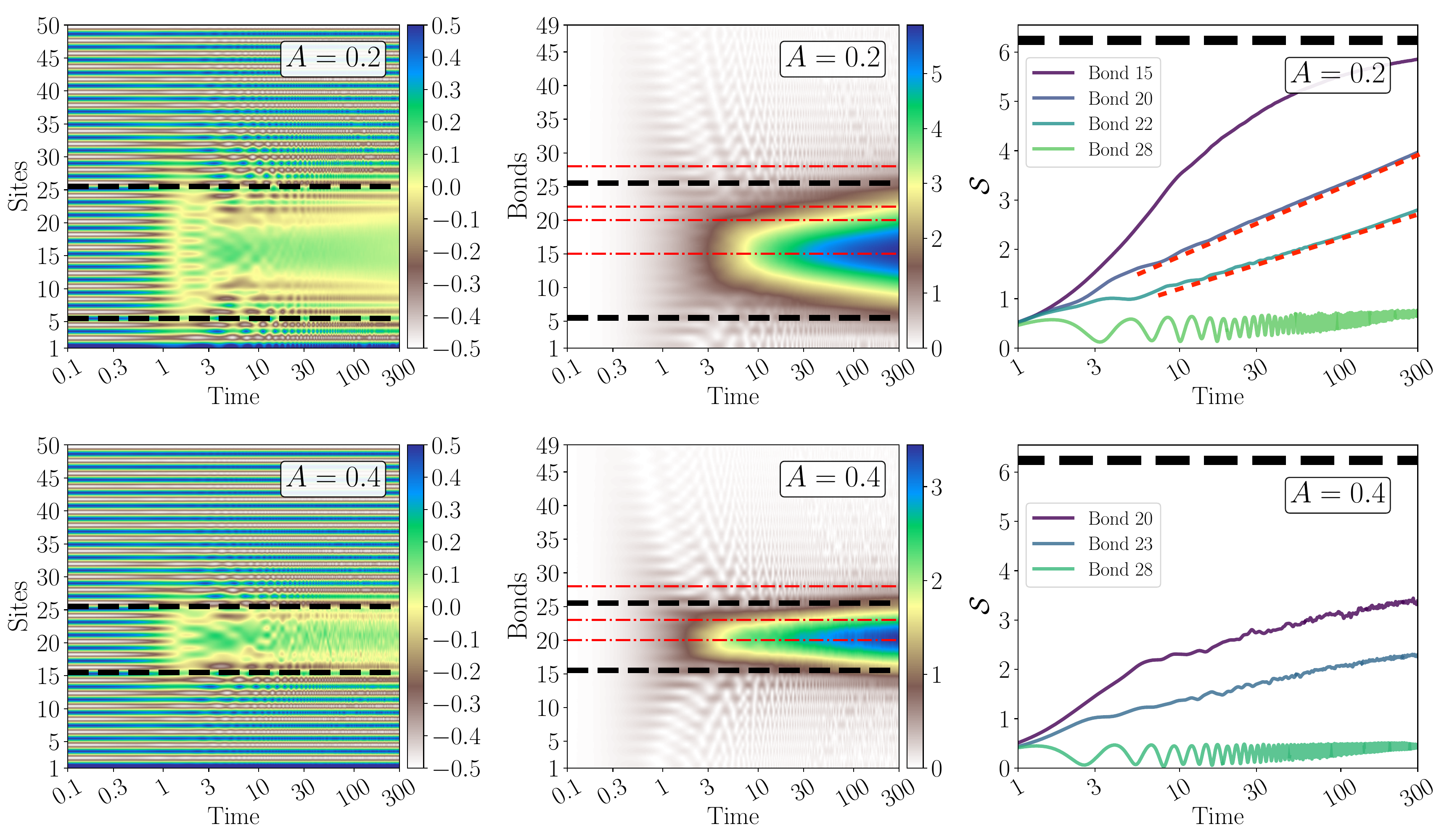}

 \caption{The Heisenberg chain with combined static field and the harmonic potential.  Here we set static field $F$ to be 2, which shifts the minima of the effective potential from the center of the system.  However, the dynamics remains exactly equivalent to the scenario of Fig. \ref{dynheis} but with shifted delocalized region. 
Other descriptions are same as in Fig. \ref{dynheis}.
\label{newtitas}
}
\end{figure}

Let us now add an additional static electric field. We expect that the delocalized region will shift to center at the position where the sum of the external field and the effective local tilt coming from the harmonic trap cancel. And indeed it is so as shown in Fig.~\ref{newtitas}.

Finally let us stress that the transition between localized and delocalized regions in our model are beyond the description introduced 
in \cite{Guardado20} and further developed in \cite{Khemani20} where it is pointed out that nonergodic dynamics may appear due to an emergent dipole conservation and Hilbert space shuttering. 
{Consider the pure tilted case, $A=0$ in \eqref{eq: XXZ} for simplicity. Indeed, if the 
global dipole ${\cal D}=F\sum_l  l \langle S^z_l(0)\rangle$ would be large initially,  the corresponding ``tilted'' energy could not be fully converted into the (extensive) kinetic energy, since the part of the Hamiltonian connected to the tilt is hyper-extensive. Thus some memory of the initial global dipole would be conserved in the system. In our case, however, we consider states from the high density of states region
(lying roughly in the middle of the spectrum)  as represented by the initial N\'eel  state that corresponds to a vanishing ${\cal D}$ and a small tilted energy.  Thus N\'eel state corresponds to a ``second class'' of states considered by  \cite{Khemani20} with extensive dipole moment that thermalize for small initial tilt. In the presence of the harmonic trap, the system becomes inhomogeneous in space which makes the arguments of \cite{Khemani20, Guardado20} hardly applicable. As discussed above, in the presence of harmonic trap, the system will always localize in the region where a local field $F=Al_0$ becomes sufficiently large (bigger than our estimate of $F_c=2$). Thus for any $A$, there exist a  critical system size such that larger systems will reveal nonergodic behavior in the regions close to edges. In some sense one may view that as a generalization of superextensive energy notion (global for a given system with a constant tilt) of \cite{Khemani20,Guardado20} to an inhomogeneous system exhibiting local tilt. It is this effective local tilt which determines the fate of the system.}

We shall see, however, that the tilted energy argument may be partially invoked for domain wall states considered below. First, however, let us discuss in some detail the spinful fermions case.

\begin{figure}
\centering
  \includegraphics[width=0.9\linewidth]{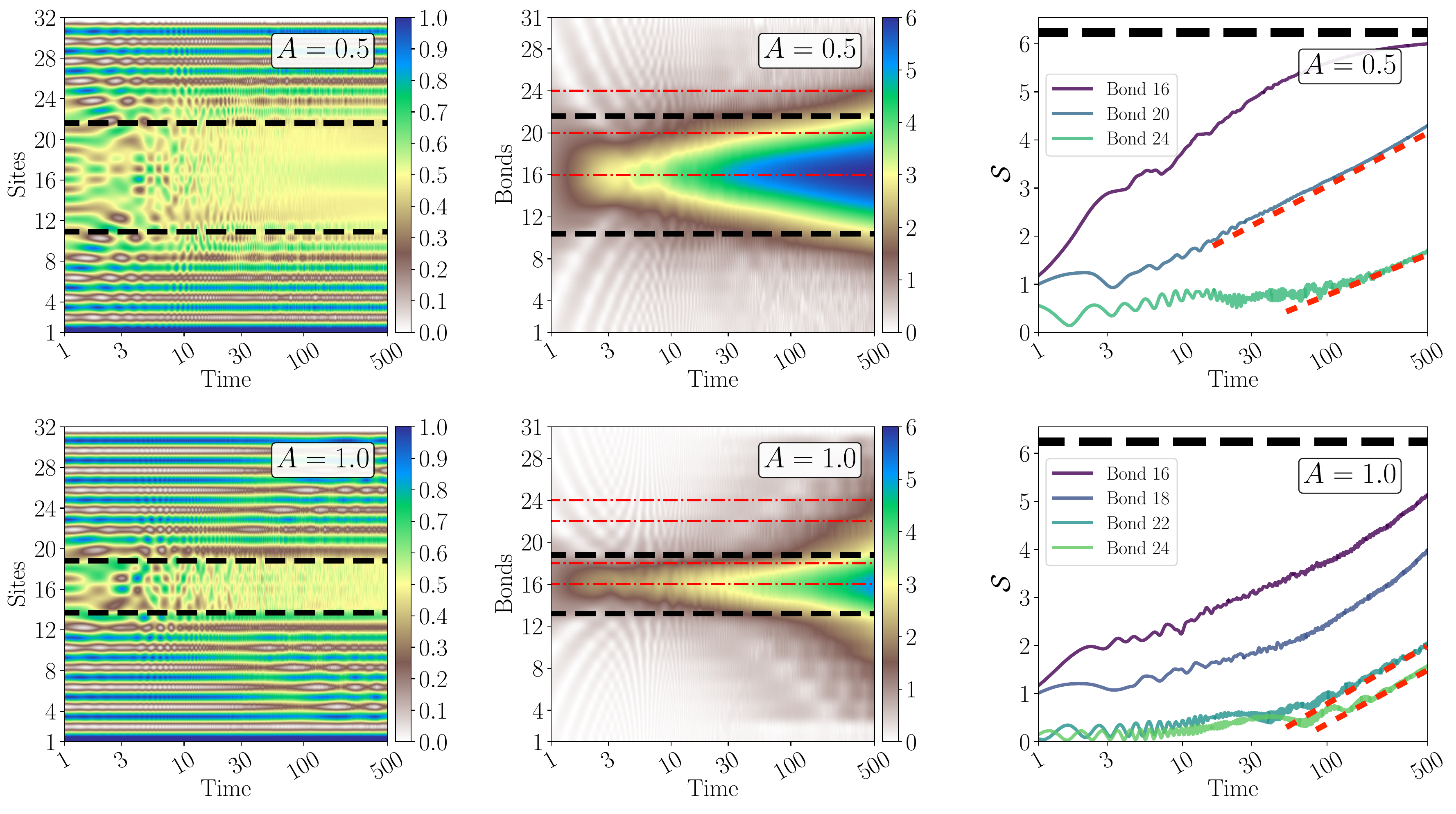}
   \caption{Left Column: Time-evolved density profile of spinful fermions for the initial staggered density-wave state under harmonic potential with no disorder.
   Middle column: Time dynamics of entanglement entropy on different bonds.
Black dashed lines gives the physical border of localization as predicted by the Stark localization with $F_c \approx 2.8$ \cite{Chanda20c}.
Right column: Entanglement entropy growth at selected bonds that are marked by red dash-dot lines in the middle column.
Red dashed lines in the right column help to compare the growth with the logarithmic trend.
    \label{l32dens}
 }
\end{figure}

\section{Spinful fermions in a harmonic trap}

Let us now discuss briefly the analogous physics for spinful fermions. Contrary to the spinless case, spinful fermions can interact on a given site making strong interactions
feasible. For that reason experimental results for disorder driven MBL was presented for spinful fermions \cite{Schreiber15,Luschen17}. Similarly one may consider
SMBL in a tilted lattice as described in \cite{Chanda20c} and recently studied experimentally \cite{Guardado20,Scherg20}. Referring the reader to these works,  let us just review here the
harmonic confinement related effects.

The Hamiltonian discussed reads:
\begin{equation}
    {H} = - J\sum_{l, \sigma} \left(\hat{c}^{\dagger}_{l \sigma} \hat{c}_{l+1 \sigma} 
		   + {\rm h.c.}\right) 
		   + U \sum_{l} \hat{n}_{l\uparrow} \hat{n}_{l\downarrow} 
		   + \sum_{l,\sigma} h_{l} \hat{n}_{l\sigma} + \frac{A}{2}\sum_{l=-L/2}^{L/2}l^2(\hat{n}_{l\uparrow}+\hat{n}_{l\downarrow} ),
\label{model}
\end{equation}
with $l \in [-L/2, L/2]$.
As before, we set $J=1$ to be the unit of energy and consider $U=1$. Note that we assume {for the time being} the same curvature $A$ for both $\uparrow,\downarrow$ spin components.

Fig.~\ref{l32dens}  shows the time dynamics of the density profile defined via $\hat{n}_{l}^{tot}=\hat{n}_{l\uparrow}+\hat{n}_{l\downarrow} $ for two different curvatures of the harmonic trap.
We show results at ``quarter filling'' with the initial separable state {having} a single fermion (with the spin alternatively pointing up and down) on odd sites and empty even sites.
 The observed dynamics has the same general features as observed for spinless fermions.
In the center of the trap  we observe a fast ``delocalization'', going away from the center we observe a transition to {apparently} localized regions closer to the left and the right edges of the chain. As before, we may define the effective local electric field as a derivative of the chemical potential.  As shown by  dashed lines in the top row of Fig.~\ref{l32dens} the local field matches   the estimate of the threshold given as  $F_c = 2.8$ given by the static field analysis \cite{Chanda20c}.

The middle and right columns in the figure show the time dynamics of the entanglement entropy across different bonds. Linear-in-time increase of the entropy in the central delocalized region changes into the slow logarithmic-like behavior in the localized parts -- that may be considered as an another evidence for localization in the outer regions.

Observe, however, that the dynamics in spinful case, as visualized in Fig.~\ref{l32dens} looks more violent. One clearly observes a nontrivial density dynamics in the localized regime, similarly while the entropy growth seems confined in space for spinless case (as seen in Fig.~\ref{dynheis}), dynamics of spinful fermions show signs of the fact that the high entropy region grows in time. While we do not have data for longer times one might speculate that for much longer times the system destabilizes. This is especially plausible if we recall that the spin dynamics in the system seems delocalized \cite{Prelovsek16,Zakrzewski18,Kozarzewski18,Sroda19} - we concentrate here on density time dynamics only.

\begin{figure}
\centering
  \includegraphics[width=0.9\linewidth]{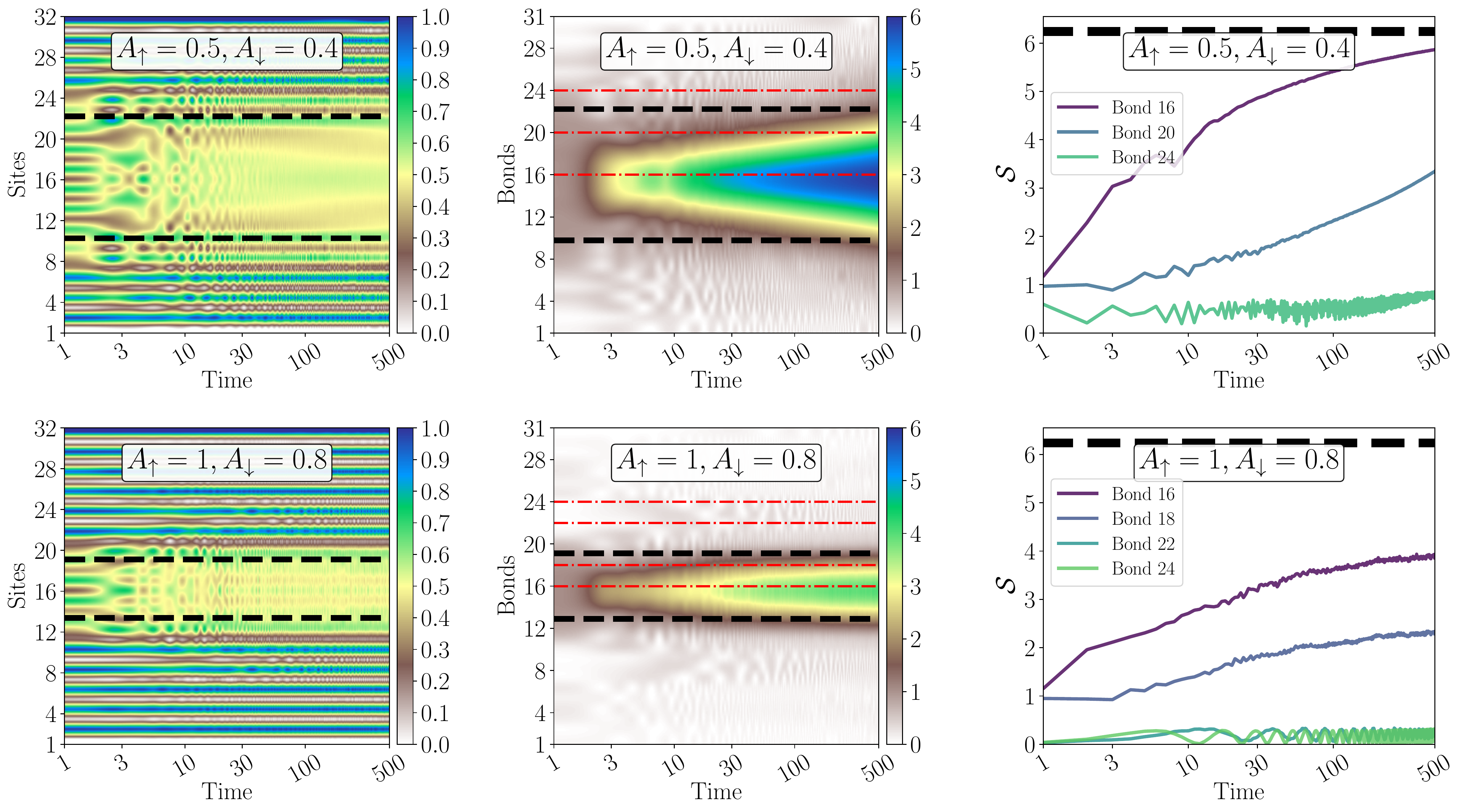}
   \caption{{Same as Fig.~\ref{l32dens} but the trap is made slightly shallower  for $\downarrow$-fermions. Observe a slower broadening of the delocalized region in time and a slower growth of the entanglement entropy as compared to $A$-symmetric case. 
    \label{l32dff}
 }}
\end{figure}
{To test whether indeed the spin diffusion is responsible for this spreading we break the symmetry between $\uparrow$-fermions and $\downarrow$-fermions by replacing the harminc trap part in \eqref{model} by the spin-dependent trap
\begin{equation}
   H_T = 
     + \frac{A_\uparrow}{2}\sum_{l=-L/2}^{L/2}l^2\hat{n}_{l\uparrow}+  \frac{A_\downarrow}{2}\sum_{l=-L/2}^{L/2}l^2\hat{n}_{l\downarrow} .
\label{mod2}
\end{equation}
That breaks the symmetry between the spin components. In such a case the standard MBL is recovered in disordered situations \cite{Kozarzewski18,Sroda19}. The resulting dynamics is presented in Fig.~\ref{l32dff}. The local tilt becomes different for different spin components so it is hard to compare the position of the crossover between localized and delocalized regions. However, while in Fig~\ref{l32dens} a careful inspection reveals that an additional weak front of delocalization may be observed (as a function of time) in
average spin dynamics (left panels) this phenomenon is entirely absent when the spin symmetry is broken by different curvatures of the potential. For broken spin symmetry the spread of the entanglement entropy fast growth region in time seems to be confined, and the growth on particular bonds weaker for broken spin symmetry despite the fact that we actually lowered one of the curvatures. Similar behavior is observed 
for weaker (top panels) and stronger (bottom panels) harmonic driving.}

{Note that for a pure tilt experiment \cite{Scherg20} the spin symmetry was also broken due to different tilt for both spin components.}

\section{Bosons in the harmonic trap}

Let us now briefly recall  the fate of the bosons in the harmonic trap \cite{Yao20b} for completeness.
The standard Bose-Hubbard Hamiltonian reads: 
\begin{equation} 
\hat H=-J\sum_{k=-L/2}^{L/2-1}(\hat{b}_k^{\dagger}\hat{b}_{k+1} + h.c.) + \frac{U}{2}\sum_{k=-L/2}^{L/2}\hat{n}_k(\hat{n}_k-1) + \sum_{k}\mu_k\hat{n}_k
\label{boson}  
\end{equation}
where $\hat{b}_k(\hat{b}_k^{\dagger})$ denote bosonic annihilator (creator)  operators obeying standard commutation relation $[\hat{b}_k, \hat{b}_l^{\dagger}] = \delta_{kl}$, $\hat{n}_k = \hat{b}_k^{\dagger}\hat{b}_k$ and $\mu_k=\frac{A}{2}(k-L/2)^2$ corresponds to the harmonic binding (for standard MBL studies \cite{Sierant18} one assumes $\mu_k$ to be random). We consider the system at half filling. Are bosons bringing some new twist to the problem?

\begin{figure}[h]
\centering
  \includegraphics[width=0.9\linewidth]{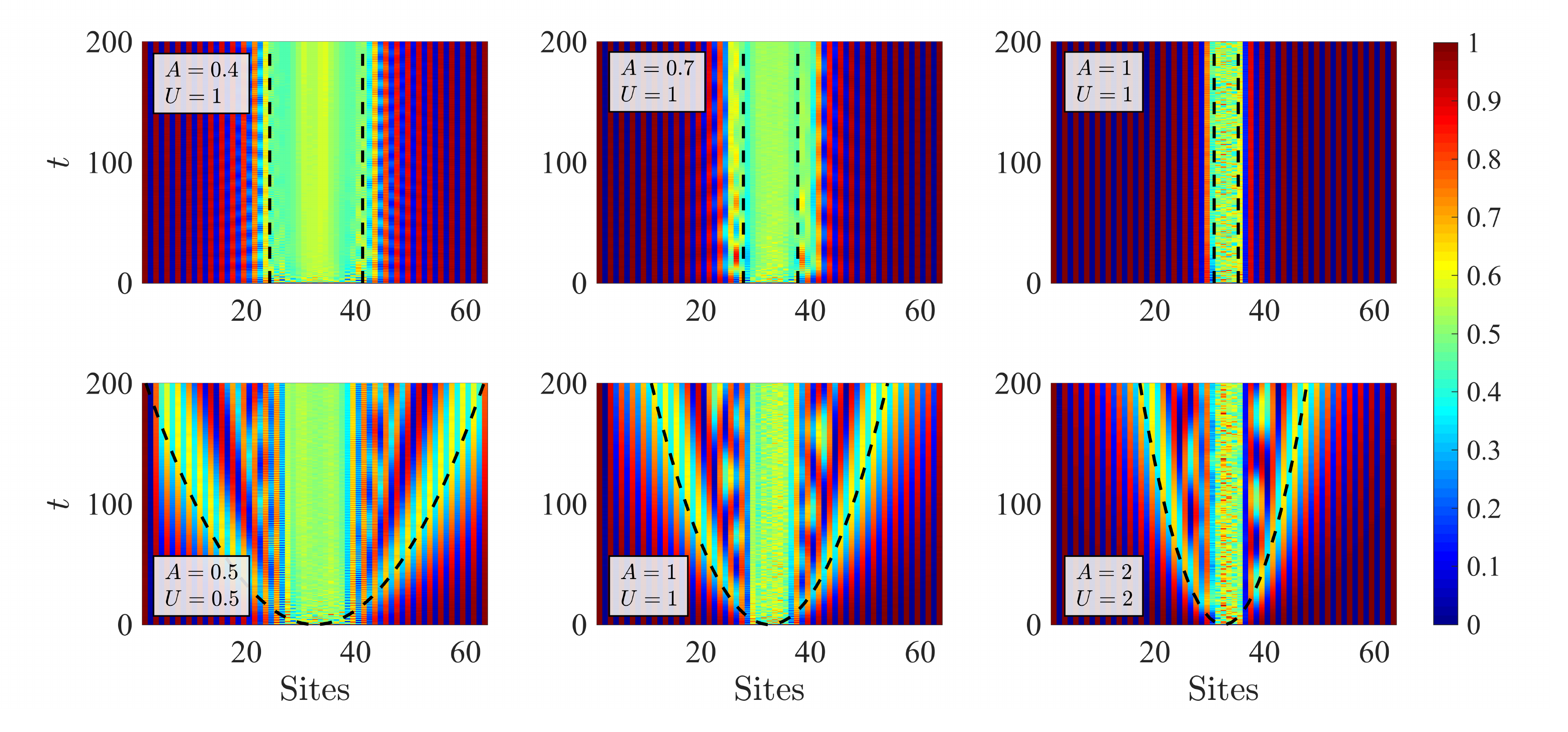}
   \caption{ Top row: Time-evolved density profiles for bosons for the initial  density-wave state under harmonic potential with no disorder for different curvatures $A$ and interaction strength $U$ as indicated in the panels.
 Black dashed lines give the border of localization as predicted by the Stark localization with $F_c \approx 3.3$ \cite{Yao20b}. Observe that the localization is partially destroyed for the resonant cases $A=U$ as shown in three bottom panels - for a discussion see text.
\label{bosonharm}
 }
\end{figure}

Fig.~\ref{bosonharm} shows the evolution of the density profiles for a number of cases considered. Again the time dynamics is accessed by TDVP algorithm, the number of bosons per site is limited to $n_{\rm max}=6$ which is largely sufficient for the average density 1/2. The initial state has a single boson on each even site with odd sites being empty
($|\psi(0)\rangle = |0,1,0,1,...\rangle$). To alert the reader we now plot the time vertically while the sites are presented horizontally, opposite to the fermionic cases discussed previously.  The top row shows a, by now, typical situation with the middle region apparently delocalized. Dashed lines give the local field estimate coming from the harmonic potential which yield  reasonable estimates of the borders between extended and localized regimes. The critical field is taken to be $F_c=3.3$ in agreement with earlier studies \cite{Sierant18,Yao20b}.

The picture is, however, markedly different for cases for which $A=U$. Apart from the delocalized region in the middle, one observes a parabolically shaped emission of particles into localized regions. This phenomenon, observed first in \cite{Yao20b} may be explained by considering the degenerate in energy subset of states
 $|\psi_j\rangle$ that are also resonant with the 
 initial state $ |\psi_0\rangle = |1,0,1,0,1,0,1,0,1,0...\rangle$:
\begin{equation}
\begin{aligned}
& |\psi_1\rangle = |0,2,0,0,1,0,1,0,1,0...\rangle \\
& |\psi_2\rangle = |1,0,0,2,0,0,1,0,1,0...\rangle \\
& |\psi_3\rangle = |1,0,1,0,0,2,0,0,1,0...\rangle \\
& \vdots
\end{aligned}
\end{equation}
The  energy difference between states $|\psi_j\rangle$ and state $|\psi_0\rangle$ is $\Delta E_{0,j} = \mu_{2j-1} + \mu_{2j+1} - 2\mu_{2j} - U = A-U$. 
Within this degenerate subspace the states are coupled by second order hopping terms with rates depending on the position. The rate dependence in $j$ 
fully explains the parabolic spreading observed in Fig.~\ref{bosonharm} - for details see \cite{Yao20b}.

\section{Domain wall physics}

\subsection{Heisenberg chain}

Localization properties may also be studied looking at domain wall melting as shown in \cite{Hauschild16}. This issue was addressed recently for the tilted Heisenberg chain
problem \cite{Doggen20s}. The authors considered a single domain wall  initial state,  $|1D\rangle$, (with initially left half of the spin pointing up and right half pointing down)  as well as
a double domain walls state, $|2D\rangle$, in which, for the system of length $L$ (with open boundary conditions) the spins $l\in[-L/4,L/4]$ are pointing up with the remaining ones pointing down.
Both initial states belong to total $S_z=0$ sector. As claimed in \cite{Doggen20s} they do not thermalize even for relatively small tilt, $F$, values  below the SMBL border as estimated in \cite{vanNieuwenburg19,Schulz19}.  This is quite interesting providing, strictly speaking, a counterexample for the eigenstate thermalization hypothesis in disorder free system. The results are obtained using TDVP for lattice size up to $L=48$. We believe that this claim has to be carefully verified analyzing different system sizes -- this is beyond
the scope of this work. Here,  rather we would like to discuss features not exposed in \cite{Doggen20s} and, in particular, consider the effect of adding a harmonic potential to the
tilted lattice.

\begin{figure}[h]
\centering
  \includegraphics[width=0.8\linewidth]{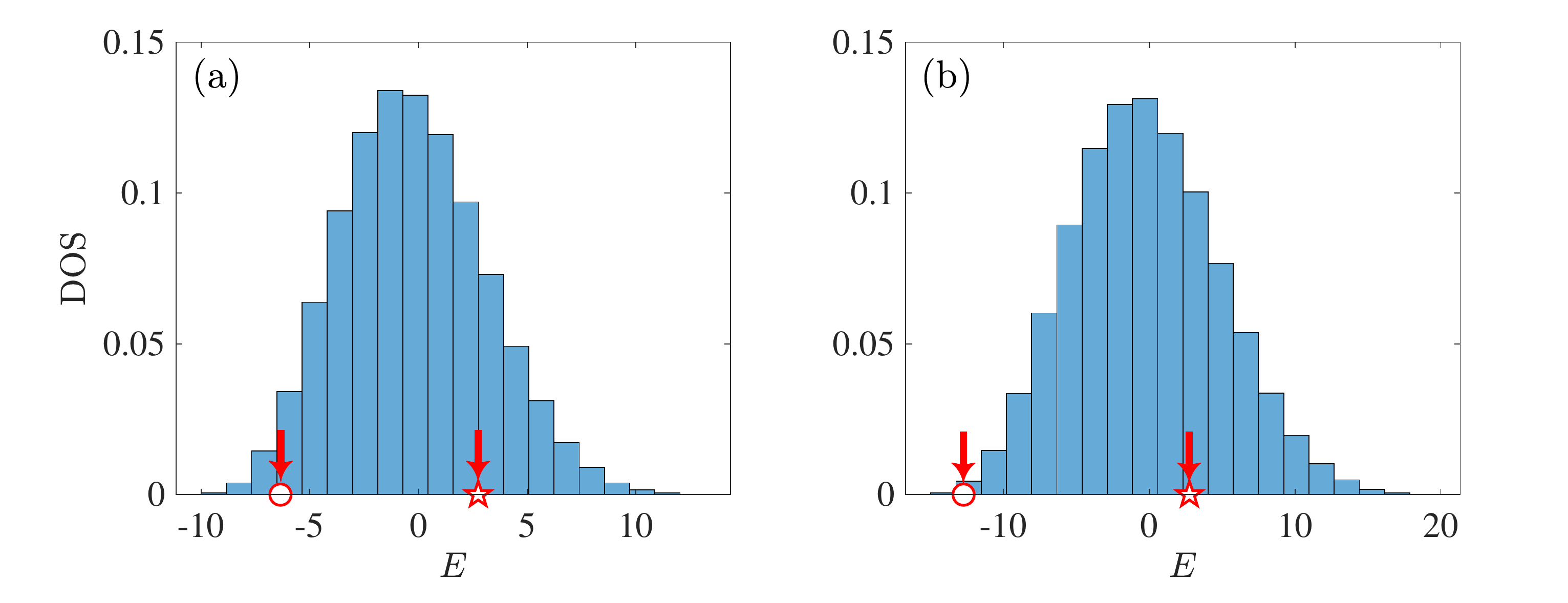}
   \caption{The density of state histogram for the Heisenberg chain with length $L = 16$, under the constant tilt  $F = 0.3$ (a) and $F = 0.5$ (b) {(with no harmonic term, $A=0$)}. The red circle denotes the position of one-domain-wall state $|1D\rangle$ in energy spectrum, and the red star denotes that of two-domain-wall state $|2D\rangle$.
    \label{hist}
 }
\end{figure}
It is important to realize where, in the energy spectrum of the system, the domain wall states are located. This is discussed qualitatively in \cite{Doggen20s} we show the example in 
Fig.~\ref{hist} where we plot the density of states (DOS) for a smaller system with $L=16$, amenable to exact diagonalization. It is apparent that the single domain wall state probes the low energy tail of DOS close to the ground state. Then it seems not so interesting for ``high temperature'' limit, i.e., the evolution in the region of maximal DOS. On the other hand the 
$|2D\rangle$ state lies close to the center of DOS being seemingly much more relevant for a generic dynamics in the system. Let us note that a close resemblance in time dynamics 
between $|1D\rangle$ and $|2D\rangle$ states were observed in \cite{Doggen20s} apparently contradicting the energy argument. 
{Let us note that \cite{Doggen20s} attributed the lack of thermalization to Hilbert space shattering mechanism \cite{Khemani20}. 

The additional explanation that we may add is that long stretches of spins pointing in the same direction in domain wall states are effectively immobile. Thus the interesting dynamic is restricted to domain walls only. Those regions carry only the portion of the energy - so the energy argument involving the full system is to some extend irrelevant. {Similarly  $|1D\rangle$ and $|2D\rangle$ states belong to different sectors of the approximate dipole  ${\cal D}$ so it is, at a first glance, not at all obvious why the dynamics of  $|1D\rangle$ and  $|2D\rangle$
states should be similar near the domain wall. If we, however, realise that  $|2D\rangle$ is in fact a combination of two  $|1D\rangle$ states practically not interacting (due to a long chain of similarly oriented spins) the Hilbert space shattering mechanism is recovered - each wall evolves (at the time scale considered) obeying a local dipole moment ${\cal D}_{local}$. The fact that the dipole moment  $|2D\rangle$ state corresponds to a different value of the global dipole moment is irrelevant. }

{Being interested in the effect of the additional harmonic field on the domain walls, }we shall consider in the following $|2D\rangle$ states only for simplicity.

\begin{figure}
\centering
  \includegraphics[width=0.9\linewidth]{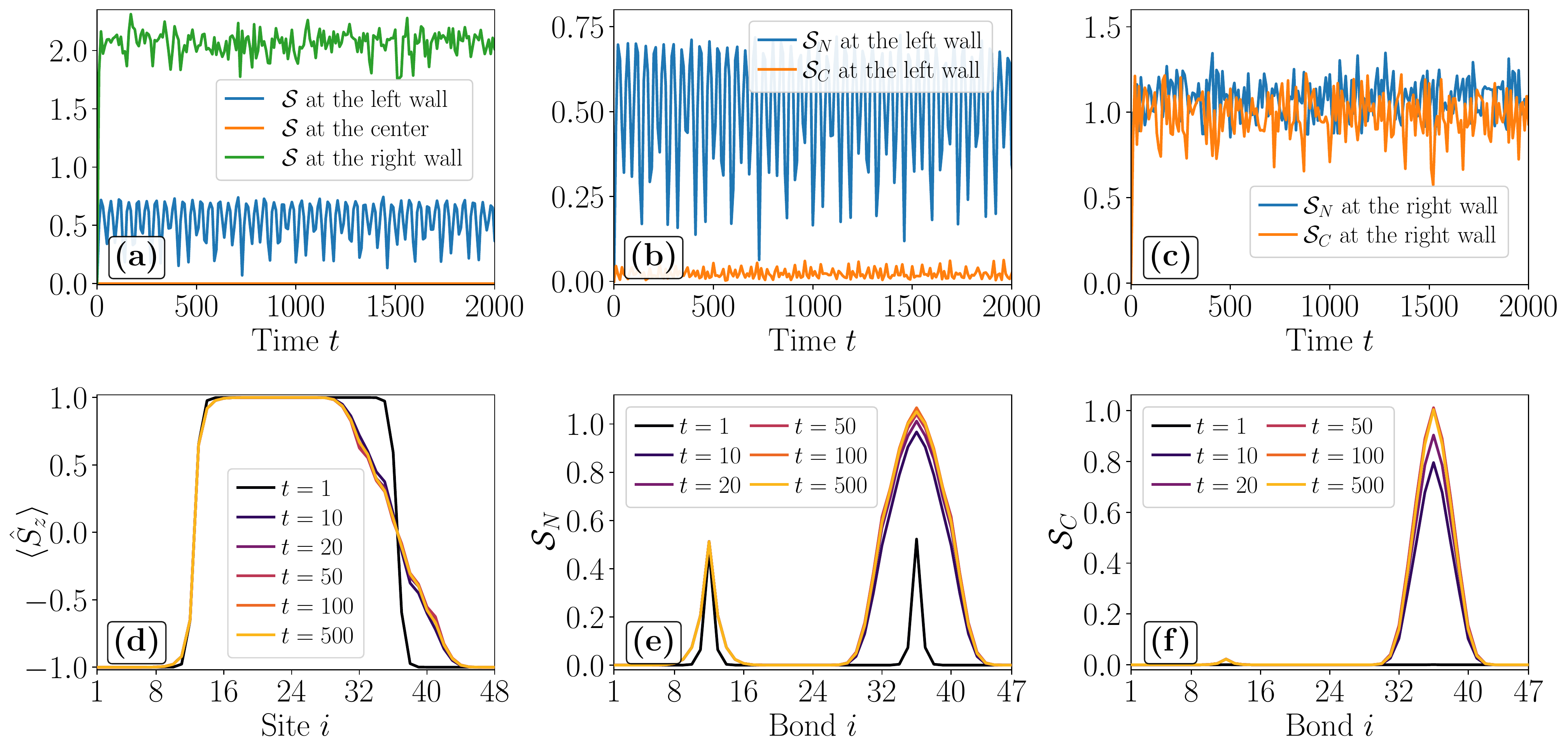}
   \caption{The time evolution of the double domain wall state {in the Heisenberg spin chain with a constant potential tilt ($F=0.5$, $A=0$).}. Top row represents the entropy dynamics. After a short initial transient the entropy saturates showing strong fast oscillations.The total entanglement entropy changes significantly at left and right wall only (a). At left wall we have mainly classical single particle tunneling while different configurations contribute to the entropy at the right well.  Averaging the fast oscillations (bottom row) reveals both the ling time sin distribution as well as    the fact that the entropy, both classical $S_N$ and quantum $S_C$, is produced mainly at the right wall. \label{spintilt}
 }
\end{figure}
While \cite{Doggen20s} consider the spin profile and the time dependence of the spin correlation function and the entropy on the domain wall we find it worthwhile to consider
enriched set of observables. {To this end we use  the entanglement entropy splitting  into two parts, a classical like ``number entropy'' and the inherently quantum ``configuration entropy'' {as introduced by \cite{Lukin19}}: 
\begin{equation} 
S(t)=- \sum_{n=0}^N p_n \log(p_n) - \sum_{n=0}^N  p_n \sum_i \rho_{ii}^{(n)} 
\log(\rho_{ii}^{(n)}) \equiv S_N(t)+S_C(t),
\end{equation}
where for $N$ particles $p_n$ is the probability to have $n$ particles in, say, left subsystem.
The number entropy $S_N$ describes a real exchange of particles between the two subsystems while $S_C$ quantifies the reshuffling of particles among different configurations in left and right part of the system. Both quantities are plotted in Fig.~\ref{spintilt}. Here we assume $F=0.5$ the value which is well below the transition to SMBL limit. The top row reveals the entropy change in time evolution. After a short initial transient the full entanglement entropy saturates at values strongly dependent on the position of the splitting.  For double domain wall state no  significant time evolution occurs in the center - the entanglement entropy is practically zero. The dynamics occurs at the right and the left domain walls, the smaller entropy at the left domain wall reveals that the dynamics there is much weaker than at the right wall. Comparison of number and configuration entropies show that at he left wall we observe a single particle transfer between the left and the right sight of the wall with no changes of other particles (as revealed by the low level of the configuration entropy). On the other hand significant configuration entropy level is reached at right wall suggesting that apart from the real particle tunneling the significant transfer between different configurations of spin occurs. The fast oscillations visible in the time domain may be removed by a high frequency filter (equivalent to an average over some range of final times).  That allows us to get the long time averages for the spin configuration as well as for the entropies presented in the bottom row.

We stress that that the entropy production is quasi-local and is concentrated to the domain wall regions. Entropy production saturates fast, reaching quasi-stationary values already at several tens of $\hbar/J$. Similarly the long time spin
profile is reached after  a few tunnelings only. {Let us note that a seemingly similar behavior was observed in \cite{Rakovszky19} for domain walls dynamics. In their situation the entropy growth is very fast having $\sqrt{t}$ character. In tilted lattice case we consider the initial growth is fast and rapidly saturates at relatively low values.}

While we cannot claim that the observed behavior persists for very long times (yet we reached maximally 10000 tunneling times without changes in the picture observed), it seems that the quasi stationary state obtained for the spin profile and entropies  reveals the equilibrium 
dynamics. If this is so that indeed, the double domain wall initial state breaks the ergodicity of the delocalized regime and avoids thermalization. This claim, however, needs, in our opinion a more detailed analysis.

\begin{figure}[h]
\centering
  \includegraphics[width=0.9\linewidth]{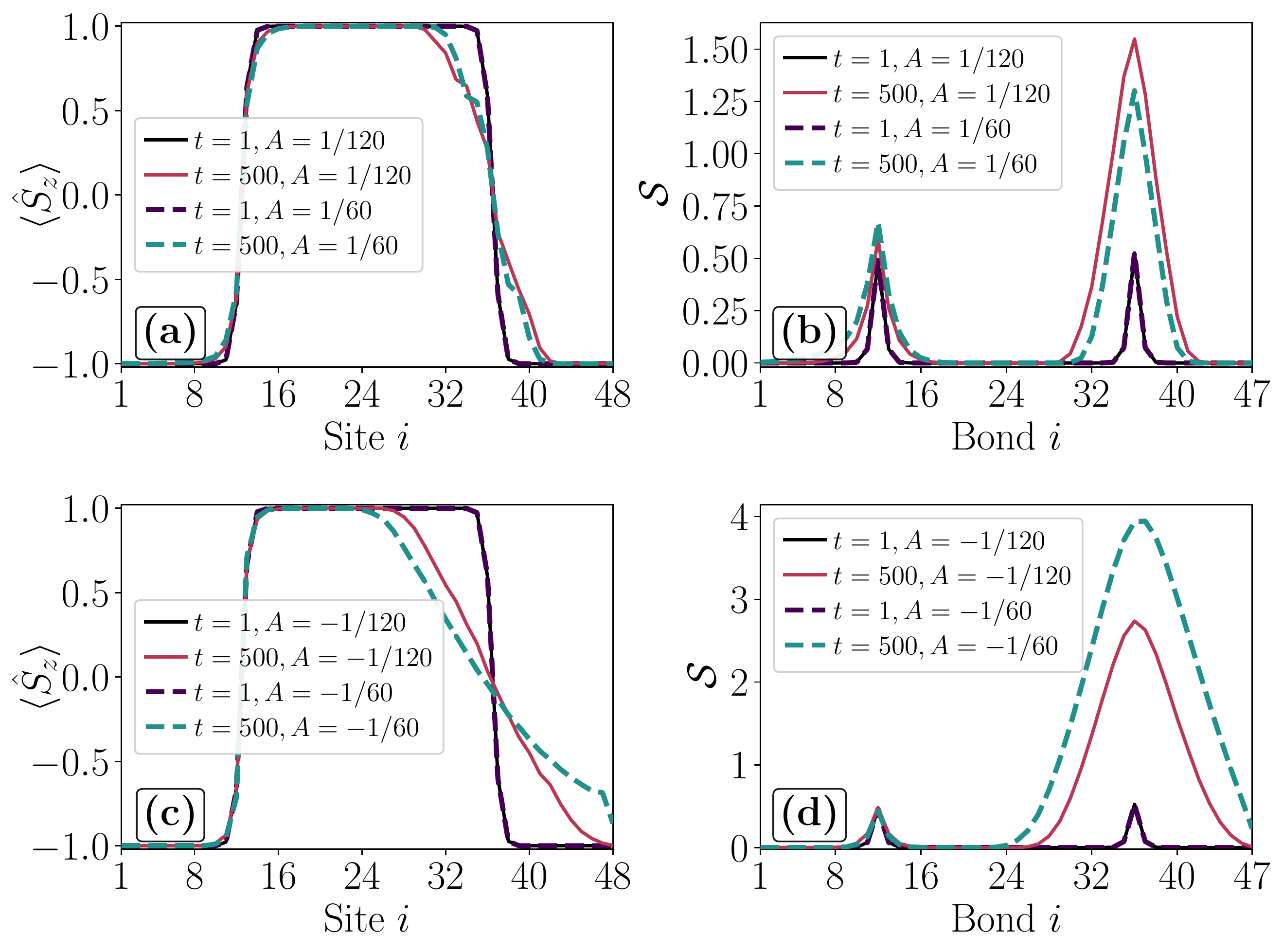}
   \caption{ The time dynamics of the double wall initial state in the combined static ($F=0.5$ as in the previous figure) and  harmonic potential. For positive harmonic curvature $A$ the effective local field adds to the static field making melting smaller than for $F=0.5$. The opposite effect occurs for negative $A$ which enhances the melting of the right wall.  -- see the discussion in the text.
    \label{spinmix}
 }
\end{figure}
Let us now add a weak harmonic confinement to the static field. The results are presented in Fig~\ref{spinmix} for two sets of values of curvature $A$. for positive $A$ the effective 
field increases at the domain wall. The  $A=1/120$ corresponds to the additional local field at the wall position of $F_{\rm loc}=0.1$ so the global field at the right wall position changes to $F_{\rm tot}=F+F_{\rm loc} =0.6$ (or 0.7 for $A=1/60$). The wall shape obtained closely resembles those obtained for solely static field case at $F=F_{\rm tot}$ indicating that
the effective local field notion works well in this case. Somewhat surprizingly, at first, it is not so for a negative curvature $A$ when local field weakens the effective field. In particular $A=-1/60$ case would lead to $F_{tot}=0.3$ at the position of the wall which for a pure static field case does not give a significant melting yet. On the other hand for the combination of the static field and a harmonic deconfinement the effect on the right wing of the domain wall is quite dramatic - the melting reaches the edge of the chain. This is due to the fact that 
the global field $F_{tot}$ decreases fast to the right of the wall as a harmonic negative confinement term becomes stronger with the distance from the center. In effect the local static field at the right edge of the system is merely $F_{tot}=0.1$. This example shows that one has to carefully apply the predictions of the local effective field induced by the harmonic trap.

Finally for this part, consider now the extreme case with no static tilt, i.e. $F=0$ and a purely harmonic strong binding. The local field is then quite large at both domain walls while becomes negligible around the center of the trap. For $A>0$ the local effective field is negative at the left and positive at the right wall. The harmonic trap acts apparently as the source of the local static field (tilt) symmetric for the left and for the right wall.

\begin{figure}[h]
\label{pure}
\centering
  \includegraphics[width=0.9\linewidth]{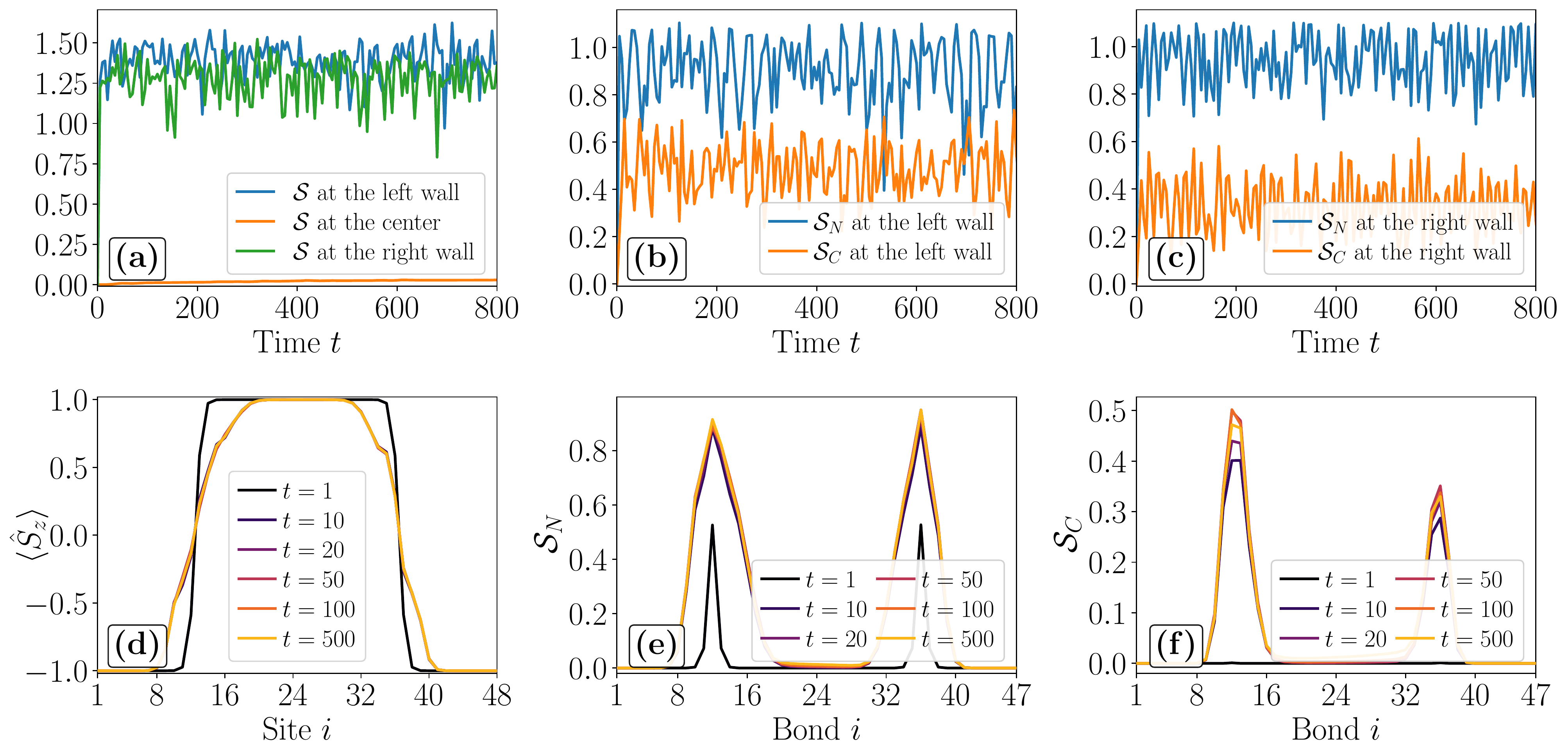}
   \caption{ The fate of the double wall initial state in the purely harmonic potential. While at the left domain wall the effective field is negative, at the right wall it is positive (with the same magnitude). The dynamics seems to be quite symmetric at both walls. $A\approx 0.058$ corresponding to  effective local fields at the positions of the walls to be $F_{\rm eff}=\pm \sim0.7$.
    \label{spinharm}
 }
\end{figure}

\subsection{Bose-Hubbard model}

Finally let us consider bosons in a combined static tilt and the harmonic potential The Hamiltonian of the system considered is:
\begin{equation}
\hat{H} = -J\sum_{\langle l,k\rangle}\hat{a}_l^{\dagger}\hat{a}_k + \frac{U}{2}\sum_l\hat{n}_l(\hat{n}_l-1) +\sum_{l=-L/2}^{L/2}(Fl+ \frac{Al^2}{2})\hat{n}_l.
\label{hmbosnew}
\end{equation}
Again the tunneling $J$ is set to be 1, and we fix interactions at $U = J$ . The time evolution is studied as, before, in unit of tunneling time $\hbar/J$. Following the dynamics with TDVP we again take a  1D lattice  of  L=48 sites. 
We consider mean half filling case with left $L/2$ states occupied by one boson initially while the right $L/2$ states being empty for a single domain wall ($|1D\rangle$) state. The double domain ($|2D\rangle$) wall state has occupied sites in $[-L/4, L/4]$ interval.  Note that in this subsection the harmonic potential is {\it not} symmetric with respect to the center of the trap, the effective local tilt is positive and adds to the constant tilt $F$ -- compare \eqref{hmbosnew}.

Time propagation for bosons is more demanding numerically as one has to restrict local Hilbert space. For the mean half feeling case we assume the  maximal atom number on site to be $n_{max} = 4$. The evolution is carried out as before restricting the auxiliary  maximal bond dimension to $\chi_{max} = 256$. In all the runs, the central bond saturates to $\chi =  \chi_{max}$ showing that the results are at the edge of the convergence. Still we believe, basing on a limited tests, that the results presented are converged to within the width of the line in the plots.

\begin{figure}[h]
\centering
\includegraphics[width=0.9\linewidth]{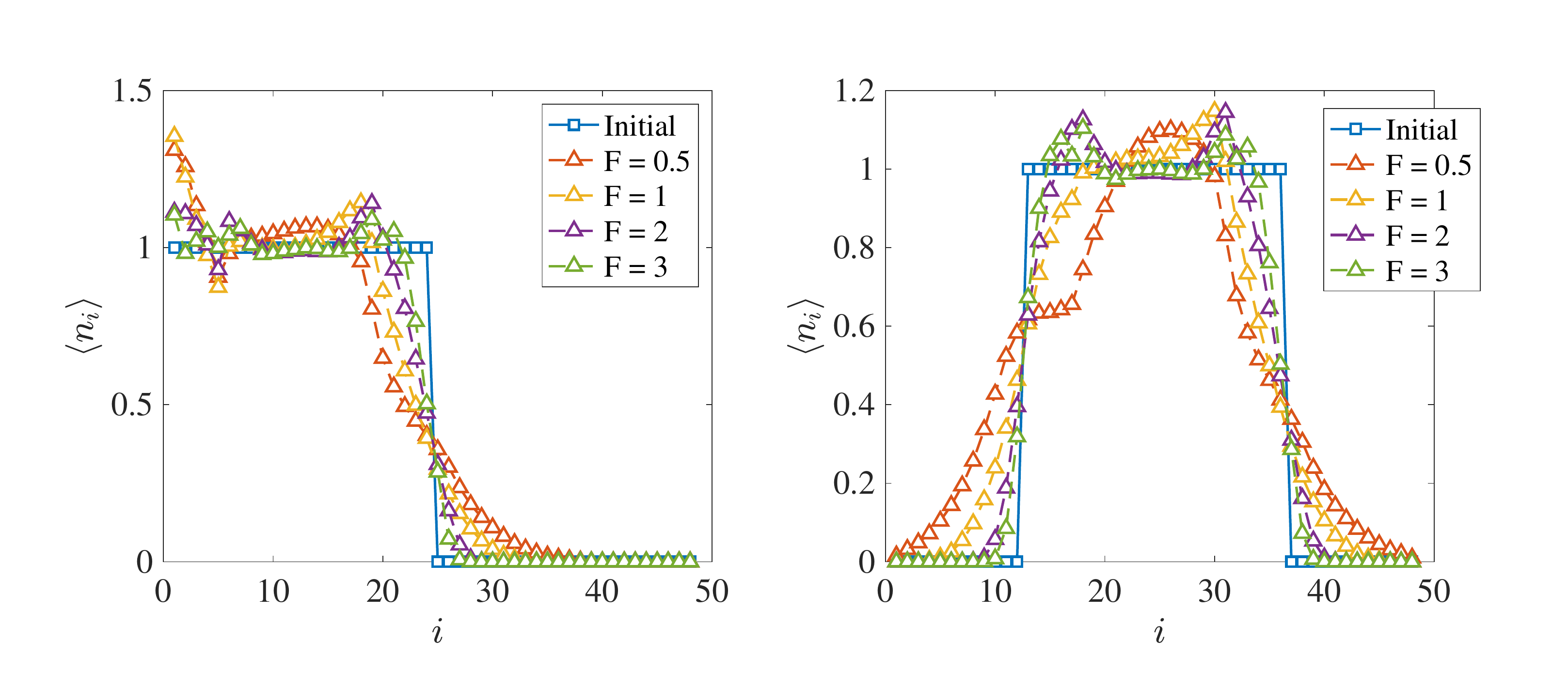}
\caption{{The occupation profile for time-evolved states under different values of the static field as indicated in the figure. The blue square denotes initial profile, $|1D\rangle$ on the left and $|2D\rangle$ on the right. The results are time averaged for $t\in[400, 500]\hbar/J$ interval (due to fast oscillations). }}
   \label{Bosprof} 
\end{figure}

Fig.~\ref{Bosprof} shows the final profiles obtained for both $|1D\rangle$ (left) and $|2D\rangle$ (right) initial states at different values of the static tilt $F$ being well below the
critical field $F_c$ estimated for the typical transition to SMBL \cite{Yao20b}. The results are averaged over a final time window to wash out the effect of rapid oscillations due
to Bloch-like oscillations. While significant deformations of initial profiles are seen resulting in the wall melting the memory of the initial state is clearly preserved up to the final times.
Note that the wall itself does not prohibit transport as in the previously discussed spinless fermion case as multiple occupation of sites is permitted. For a single domain wall  a slight accumulation of particles in visible in the left part of the lattice. Similar excess occupation is visible for the $|2D\rangle$ in the middle of the chain

The melting dynamics may be put on a bit more quantitative basis by calculating the final melting range $\xi$. We define it as the distance between the site at which the initial population changes from unity to 0.95 and the position (on the other side of the domain  wall) where the population changes from 0 by 0.05. The resulting $\xi(F)$ dependence is shown
in Fig.~\ref{melt}. Observe that for both single and double domain states two sets of points overlap with very good accuracy. The blue squares correspond to a pure linear tilt with a given $F$ value. The red symbols are obtained  at field values corrected by the local tilt $\Delta F = A l_d$ due to the presence of an additional harmonic potential
with $AL^2=10$ -- compare \eqref{hmbosnew}. The excellent agreement shows that the local field approximation works very well for bosons.

\begin{figure}[h]
\centering
\includegraphics[width=0.9\linewidth]{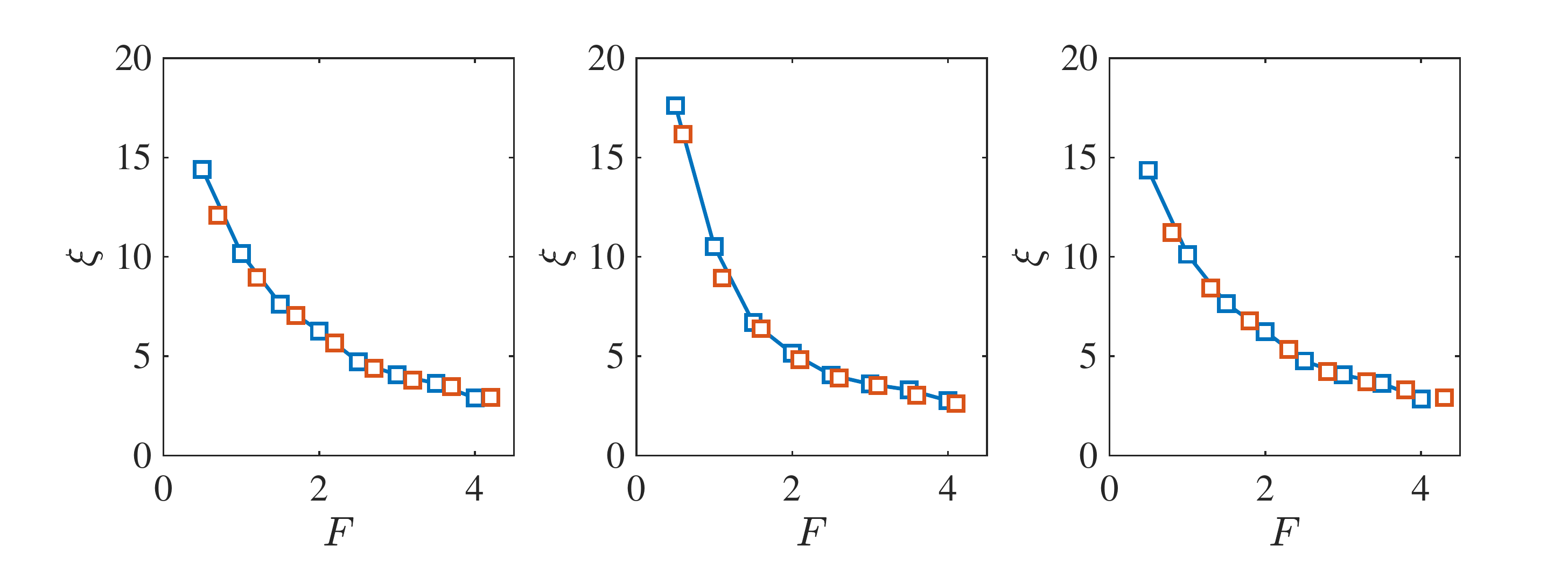}
\caption{{The melting range $\xi$ calculated from different domain wall: (a) The central domain wall of $|1D\rangle$  states, (b) The left domain wall of $|2D\rangle$  state, and (c) The right domain wall of $|2D\rangle$  state. The blue squares denote ranges calculated for a pure static force case while the red squares are for $AL^2 = 10$ harmonic trap with a horizontal shift $\Delta F$. The $\Delta F$ is determined by the domain wall position $l_d$: $\Delta F = Al_d$ and the collapse of different data points shows the good performance of local field approximation.}}
   \label{melt} 
\end{figure}

\section{Conclusions}

We discussed the time dynamics of 1D lattice systems in a disorder-free potential combing the static field (a constant in space tilt of the lattice) with the harmonic potential. Realistic lattice sizes of the order of $L=50$ were studied using TDVP numerical routines. In particular we reviewed the case of pure harmonic binding that may lead to the phase separation and the coexistence of extended and localized regions. The border between the two regions can be well estimated using the notion of the local electric field obtained from the derivative of the harmonic potential -- such an approach works quite well for sufficiently large lattices. This may be understood by the fact that the localization, when it sets in the many particle system has a short localization length of few physical sites only.

 We considered spinless or (briefly) spinful fermions, as well as bosonic systems. Manipulating with static and harmonic trap one can shift the  localized regions as presented for spinless fermions. We have noted that the presence of the localized regions slows down considerably the entanglement entropy growth also on the delocalized size.

We have studied in detail in combined static and harmonic potential the dynamics of  domain wall initial states. Their very slow and apparently nonergodic (on a relatively short time scale of few thousands of the tunneling times studied) dynamics is related to the stretches of spins pointing in the same direction - such a situation prevents any real transport for spin 1/2 particles. In effect the growth of entropy (being it the quantum one due to configuration change or the number entropy due to real particle transport over a studied bond) is limited to the very vicinity of walls separating domains (for spins). When no ``frozen spin'' stretches occurs, as for bosons, the melting occurs more effectively although it is still a quite slow process and the memory of the initial state shape persists on the time scale of few hundreds of tunneling times. Again the local effective field due to a derivative of the harmonic potential describes very well the physics studied. It suggests that such a local field approach may work quite well for arbitrary smooth potentials.

\section*{Acknowledgments} 
J.Z.  acknowledges interesting discussions with Piotr Sierant.
The numerical computations have been possible thanks to  High-Performance Computing Platform of Peking University as well as PL-Grid Infrastructure.
The TDVP simulations have been performed using ITensor library (\url{https://itensor.org}).
This research has been supported by 
National Science Centre (Poland) under projects  2017/25/Z/ST2/03029 (T.C.) and  2019/35/B/ST2/00034 (J.Z.)

\section*{References}

\bibliographystyle{apsrev4-1}

\end{document}